\pdfoutput=1
\pdfoutput=1
\pdfoutput=1
\pdfoutput=1
\pdfoutput=1
\documentclass[final,5p,times,twocolumn]{elsarticle}

\usepackage{amssymb}
\usepackage{color}
\usepackage{amsmath}

\usepackage{multirow}
\usepackage{booktabs}
\usepackage{bm}
\usepackage{braket}
\usepackage{amsmath}
\usepackage{graphicx}
\usepackage {ulem}
\biboptions{sort&compress}

\usepackage[pdfstartview=FitH, colorlinks,
 linkcolor={red!60!black!},
 anchorcolor={green!80!black!},
 urlcolor={blue!80!black!},
 citecolor={pink!50!black!}]{hyperref}
\usepackage[table]{xcolor}

\usepackage[T1]{fontenc}
\usepackage{lmodern} 

\journal{Physics Letters B}

\begin{document}


\begin{frontmatter}




\title{Gamow shell model predictions for six-proton unbound nucleus $^{20}$Si}


\author[ad1,ad2]{J.L. Wang}
\author[ad1,ad2]{M.R. Xie}
\author[ad4,ad1]{K.H. Li}
\author[ad5,ad1]{P.Y. Wang}
\author[ad1,ad2]{N. Michel}
\author[ad1,ad2]{Q. Yuan}
\author[ad1,ad2,ad3]{J.G. Li \corref{correspondence}}

\address[ad1]{State Key Laboratory of Heavy Ion Science and Technology, Institute of Modern Physics, Chinese
Academy of Sciences, Lanzhou 730000, China}

\address[ad2]{School of Nuclear Science and Technology, University of Chinese Academy of Sciences, Beijing 100049, China}

\address[ad3]{Southern Center for Nuclear-Science Theory (SCNT), Institute of Modern Physics, Chinese Academy of Sciences, Huizhou 516000, China}

\address[ad4]{College of Physics, Henan Normal University, Xinxiang 453007, China}

\address[ad5]{Advanced Energy Science and Technology Guangdong Laboratory, 516007, Huizhou, China}

\cortext[correspondence]{Corresponding author:  jianguo\_li@impcas.ac.cn (J.G. Li)}

\begin{abstract}

Proton-rich nuclei beyond the proton drip line are of great interest in nuclear structure physics, due to exotic phenomena such as proton emissions and the Thomas-Ehrman shift (TES). In this work, we employ the Gamow shell model (GSM) to investigate the structure and decay of $^{20}$Si, a candidate for six-proton (6$p$) emission, which can be produced via two-neutron knockout from the drip line nucleus $^{22}$Si. We predict that its ground state decays via 6$p$ emission to the ground state of $^{14}$O, with a decay energy $E_{6p} = 10.125$ MeV and a width of 371~keV. A $2^+$ state is predicted at 1.7 MeV, comparable with that in $^{18}$Mg, indicating the disappearance of the $Z=14$ magic number in $^{20}$Si. Instead, analyses of the many-body configurations and the average occupancies of the mirror states suggest the presence of $dynamic$ TES in low-lying states of $^{19}$Al/$^{19}$C and $^{20}$Si/$^{20}$C. Further evidence is provided by analyzing the contributions of different components of the GSM Hamiltonian. Moreover, this study offers the first theoretical description of $^{20}$Si and guidance for future experiments.

\end{abstract}


\end{frontmatter}

\section{Introduction}

With advances in experimental instrumentation and techniques, more and more proton-rich nuclei near (and beyond) the proton drip line have been observed, along with a variety of exotic structures and dynamics, including cluster
structures~\cite{RevModPhys.90.035004,PhysRevLett.112.162501,10.1143/PTPS.192.1,PhysRevC.85.054320}, shell evolution~\cite{SORLIN2008602,Ye2025}, Thomas-Ehrman shift (TES)~\cite{PhysRev.88.1109,PhysRev.81.412,OGAWA1999157}, and proton emission~\cite{PhysRevC.78.041302,PhysRevC.85.034327,PhysRevC.89.044610,PhysRevC.102.044614,PhysRevLett.108.142503}. 
Among these, proton emission emerges as a primary focus of significant interest.
Experimentally, the properties of proton-rich nuclei beyond drip line are typically deduced from decay-product analyses, employing invariant mass spectra or angular correlation measurements~\cite{PhysRevLett.131.172501,PhysRevLett.127.262502,hkmy-yfdk,PhysRevC.110.L031301}.

Various proton-emission modes have been experimentally discovered.
Owing to pairing-induced odd-even staggering of the drip line, one-proton ($1p$) emission typically dominates in odd-$Z$ isotopes, whereas two-proton ($2p$) emission is favored in even-$Z$ systems~\cite{RevModPhys.84.567,PhysRevC.101.014319}.
For example, $1p$ decay of $^{15}$F~\cite{PhysRevC.69.031302} and $2p$ decay of $^{6}$Be~\cite{PhysRevLett.109.202502}, $^{12}$O~\cite{PhysRevLett.74.860}, and $^{16}$Ne
~\cite{PhysRevLett.113.232501}. 
Even more exotic multiproton modes have been observed further beyond the drip line:
$3p$ emitters $^{7}$B~\cite{PhysRevC.84.014320}, $^{13}$F~\cite{PhysRevLett.126.132501}, $^{17}$Na~\cite{PhysRevC.95.044326}, $^{20}$Al~\cite{hkmy-yfdk} and $^{31}$K~\cite{PhysRevLett.123.092502} show a sequential 1$p$-2$p$ decay mechanism, while the $4p$ emitters $^{8}$C~\cite{PhysRevC.82.041304} and $^{18}$Mg~\cite{PhysRevLett.127.262502}, decaying via sequential 2$p$-2$p$ emissions, have also been experimentally observed.
Remarkably, the extremely unstable nucleus $^{9}$N has been experimentally confirmed as the first and only observed ground-state $5p$ emitter, providing valuable insight into multiproton emission mechanisms~\cite{PhysRevLett.131.172501}.
Moreover, it is established that the nuclei $^{15}\text{F}$, $^{16}\text{Ne}$, $^{17}\text{Na}$, and $^{18}\text{Mg}$ decay via single- or multi-proton emission to the $^{14}$O ground state~\cite{PhysRevC.69.031302,PhysRevLett.113.232501,PhysRevC.95.044326,PhysRevLett.127.262502}. $^{20}$Si, composed of $^{14}$O plus six protons, emerges as a popular candidate for 6$p$ emission.
The $4p$ emitter $^{18}$Mg was produced via $2n$ knockout reactions from a $^{20}$Mg beam impinging on a $^{9}$Be target~\cite{PhysRevLett.127.262502}. By analogy, $^{20}\text{Si}$ is expected to be produced via $2n$ knockout from a $^{22}\text{Si}$ beam impinging on a $^{9}\text{Be}$ target. This prospect is particularly relevant since $^{22}\text{Si}$ has recently been identified as a bound nucleus and fixes the proton dripline location for the Si element~\cite{ffwt-n7yc}.

Independent lines of evidence also indicate pronounced isospin-symmetry breaking in proton-rich emitters and their mirror partners~\cite{PhysRevLett.110.172505,BACZYK2018178,PhysRevC.104.014324,PhysRevC.66.024314,RevModPhys.94.025007}. While isospin symmetry implies that mirror nuclei should exhibit closely analogous spectra~\cite{Heisenberg1932,PhysRev.51.106,BENTLEY2007497}, large mirror-energy differences, often attributed to the TES, are frequently observed between analog states~\cite{hkmy-yfdk,PhysRevLett.123.142501,PhysRevLett.122.192502,Hoff2020,Zhou2024,PhysRevC.111.034327}. TES is strongly influenced by continuum coupling and the associated configuration mixing~\cite{PhysRevC.65.051302}.
Beyond the $static$ TES picture, a configuration-driven $dynamic$ TES has been proposed: when mirror partners differ in their dominant many-body structures, the resulting asymmetry can amplify TES effects~\cite{PhysRevLett.88.042502,PhysRevC.91.024325,PhysRevC.100.024304,PhysRevC.101.034312}. $Dynamic$ TES has been documented, for example, in the mirror pairs $^{16}$Ne/$^{16}$C and $^{18}$Mg/$^{18}$C~\cite{PhysRevC.91.024325,PhysRevC.111.014302}.

Accurately describing proton-rich nuclei requires a simultaneous treatment of internucleon correlations and coupling continuum, which poses significant challenges for reliable theoretical studies~\cite{DOBACZEWSKI2007432}. The Gamow shell model (GSM) provides a unified framework that treats correlations and continuum coupling~\cite{Michel_2009,Michel_Springer,PhysRevC.88.044318,physics3040062} and has been successfully applied to a range of proton-rich systems~\cite{PhysRevC.111.014302}.
For example, the possible two-proton radioactivity of $^{38,39}$Ti~\cite{HUANG2025139257}, isospin symmetry breaking in $^{20}$Al/$^{20}$N~\cite{hkmy-yfdk}, and the new magic numbers $Z=14$ in $^{22}$Si~\cite{ffwt-n7yc}.
In this study, we employ GSM to make predictions for the $6p$ unbound nucleus $^{20}$Si and provide valuable guidance for future experimental studies.

In this paper, we first introduce the theoretical framework of the GSM. 
We assess the feasibility of detecting $^{20}$Si through 6$p$ emission by systematically examining the excitation energies and widths of the ground and excited states in $^{14}$O isotones of $A = 15\text{-}18$. 
After this, we analyze the $dynamic$ TES properties in $^{19}$Al/$^{19}$C and $^{20}$Si/$^{20}$C. The many-body configurations and the average occupancies of the $s_{1/2}$ and $d_{5/2}$ partial waves are then discussed. To further elucidate the mechanisms of TES, we also calculate the contributions of
different components of the GSM Hamiltonian. Finally, reliable conclusions are made.

\section{Method}

GSM is a multiconfiguration interaction framework with a core
plus valence nucleon(s) picture~\cite{PhysRevLett.89.042501,PhysRevLett.89.042502,Michel_2009}. It originates from an extension of the traditional shell model (SM) into
the complex-energy plane by replacing the harmonic oscillator (HO) basis with the Berggren basis. The Berggren basis consists of one-body states, including bound, resonance, and complex-energy scattering states, generated by a finite-range potential~\cite{PhysRevC.47.768,BERGGREN1968265}. The Berggren completeness of the Berggren basis reads:
\begin{equation}\label{Eq1}
    \sum_n |u_n\rangle \langle u_n| + \int_{L^+} |u(k)\rangle \langle u(k)| dk=\mathbf{\hat{1}},
\end{equation}
where $|u_n\rangle$ is the bound or resonant one-body state, and $|u(k)\rangle$ is the scattering state belonging to the $L_+$ contour of complex momenta. To use the Berggren basis in practical numerical applications, the contour $L_+$ is discretized with a Gauss-Legendre quadrature to obtain an eigenproblem similar to that of HO-SM~\cite{Michel_2009}. One typically needs 30–50 discretized states in order to have converged results~\cite{PhysRevC.83.034325}. 
Then, the Slater determinants built from the one-body states of the Berggren completeness relation serve as the many-body basis for the GSM, within which the Hamiltonian is diagonalized to a complex symmetric matrix.

To avoid spurious Center-of-mass (CM) excitations in GSM
wave functions, the separation between internal and CM motion is achieved within the cluster orbital shell model (COSM) framework, where all coordinates are defined relative to the chosen inner core CM picture~\cite{PhysRevC.38.410}. The resulting GSM Hamiltonian takes the form,
\begin{equation}\label{Eq2}
    H=\sum_{i=1}^{A_{\textup{val}}}(\frac{\mathbf{\hat{p}}_{i}^{2}}{2\mu_{i}}+\hat{U}_i)+\sum_{i< j\in \textup{val}}(\hat{V}_{ij}+\frac{\mathbf{\hat{p}}_i \cdot \mathbf{\hat{p}}_j}{M_{\textup{core}}}),
\end{equation}
where $A_{\textup{val}}$ denotes the number of valence nucleons, 
$\mu_i$ is the reduced mass of the nucleon, and $\hat{U}_i$ represents the core-nucleon potential, modeled using a one-body Woods-Saxon potential.
The residual internucleon interaction $\hat{V}_{ij}$ is based on the two-body part of pionless effective field theory at next-to-leading order.
The missing effects of three-body interactions are simulated by introducing an $A$-dependent one-body factor $F_{1b}$ and two-body factor $F_{2b}$ to the respective one-body and two-body parts of the Hamiltonian.
In addition, the Coulomb interaction is explicitly included for valence protons. $M_{\textup{core}}$ is the mass of the core, and the term $\mathbf{\hat{p}}_i \cdot \mathbf{\hat{p}}_j/M_{\textup{core}}$ takes into account the recoil of the active nucleons.
We adopt the same parameters for the WS core potential as in Ref.~\cite{PhysRevC.103.044319}.
The parameters for the residual nucleon-nucleon interaction and the $A$-dependent one- and two-body factor are the same as in Ref.~\cite{PhysRevC.100.064303,PhysRevC.103.044319}.

Following the GSM calculations in Ref.~\cite{PhysRevC.100.064303,PhysRevC.103.044319}, we take into account $spdf$ partial waves in the valence space (proton type or neutron type). Partial waves of the $spd$ are represented using the Berggren basis, whereas the HO basis is used for the $f$ partial waves whose effect on asymptotic many-body wave functions is negligible because of the large centrifugal barrier.
In the frame of the core + valence nucleon picture, we use a $^{14}$C core for the carbon isotopes, while for the carbon isotones, a $^{14}$O core is adopted.

The GSM Hamiltonian matrix is diagonalized using the Jacobi-Davidson method~\cite{MICHEL2020106978}, combined with the overlap method~\cite{PhysRevC.67.054311,Michel_2009}. However, an excessively large model space leads to a high dimension of the Hamiltonian. We first perform an initial GSM calculation within the Berggren basis by allowing at most two scattering states to be occupied in the continuum, denoted as 2p-2h truncations, where the one-body density matrix from the GSM Hamiltonian is diagonalized to optimize the single-particle basis into natural orbitals~\cite{Brillouin1933}. The natural orbitals recapture a large part of the strength of many-body systems, and the GSM model space dimension could be much reduced. On this optimized basis, we then conduct the GSM calculation with the 2p-2h truncation, whereby convergence is well obtained. Owing to the properties of the complex Hamiltonian, the computed eigenenergies are complex quantities, in which the real parts correspond to the calculated energies of the atomic nucleus, and the widths can be determined from the imaginary parts.

\section{Results}

\begin{figure}[htbp]
    \centering
    \includegraphics[width=0.48\textwidth]{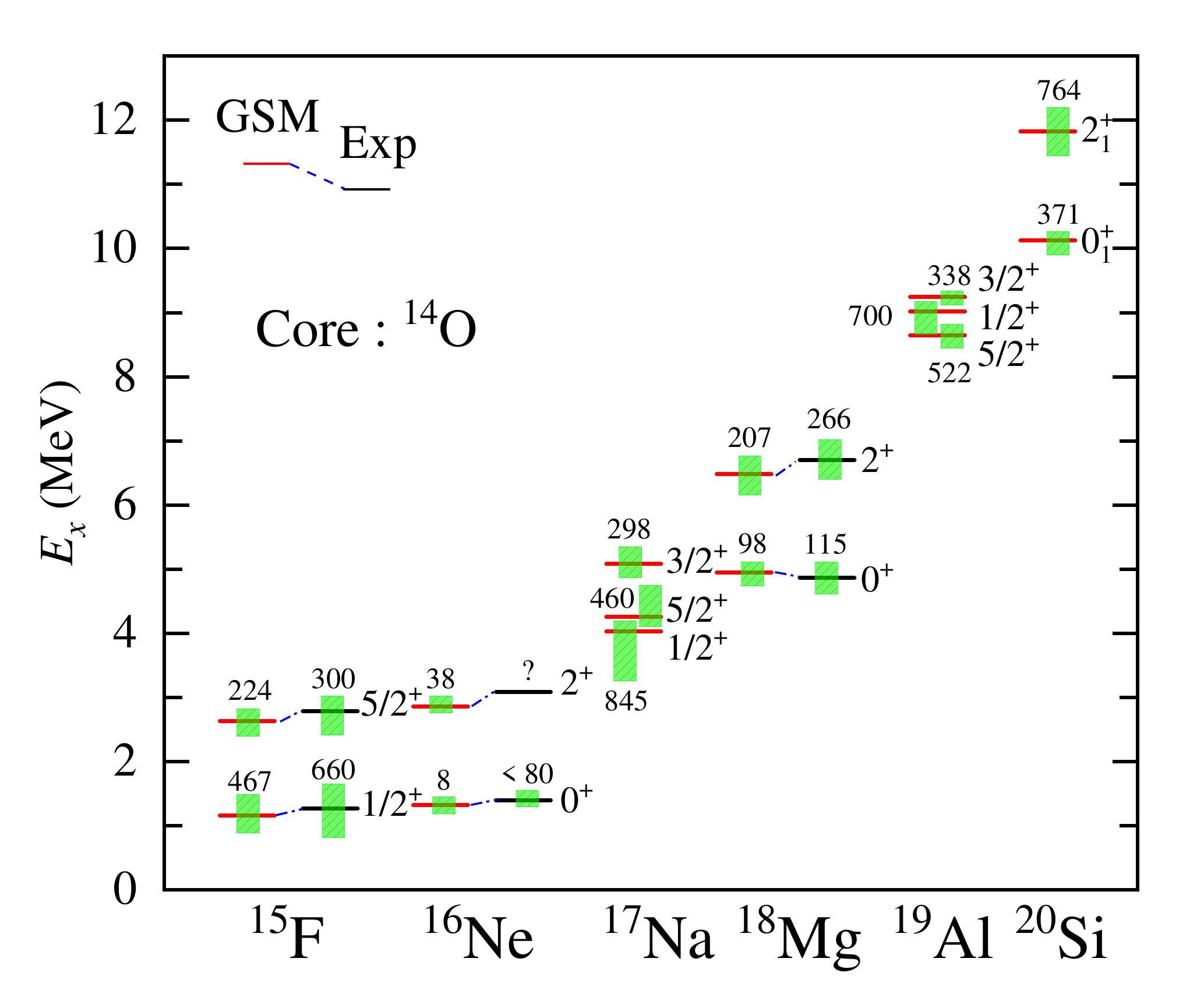}
\caption{The calculated energies ($E_{x}$, in MeV) and widths (in keV) for ground and excited states of carbon isotones with GSM relative to the $\rm{^{14}O}$ core, along with available experimental data. Green striped squares represent the widths, and the corresponding values are written nearby. Experimental data for $\rm{^{15}F}$, $\rm{^{16}Ne}$ and $\rm{^{18}Mg}$ are obtained from Ref.~\cite{ensdf,PhysRevLett.113.232501,PhysRevLett.127.262502}.}
    \label{absolute energy}
\end{figure}

The $^{14}$O isotones with mass numbers $ A= 15\text{-}18 $, located beyond the proton drip line, are established proton emitters that decay by emitting one or more protons to the $^{14}$O ground state. We performed GSM calculations for these isotones; the resulting energies and widths of the low-lying states are shown in Fig.~\ref{absolute energy}, along with available experimental data~\cite{ensdf,PhysRevLett.113.232501,PhysRevLett.127.262502}. Overall, the experimental systematics are well reproduced.
To estimate the uncertainties, we systematically examined the sensitivity of the results to the model space, the truncation scheme, and the parameter of the EFT interaction, focusing on the low-lying states of $^{18}$Mg, $^{19}$Al, and $^{20}$Si. First, we extended the GSM calculations by including the $g_{7/2}$ and $g_{9/2}$ partial waves with the fixed parameters for GSM Hamiltonian, which cause negligible changes: the energy and width shifts are about 50~keV and less than 100~keV, respectively, while for $^{18}$Mg, both shifts remain below 5~keV for $^{19}$Al and $^{20}$Si. Consequently, $g_{7/2}$ and $g_{9/2}$ partial waves were omitted in subsequent analyses.
To assess truncation effects, we further extended the GSM calculations within a natural orbital basis from the 2p-2h to 3p-3h configurations. In the case of $^{18}$Mg, the 3p-3h results are well converged, with differences within 1~keV, and the results from both truncation schemes agree with experimental data~\cite{PhysRevLett.127.262502} and Ref.~\cite{PhysRevC.111.014302}. 
For $^{19}$Al and $^{20}$Si, however, the 3p-3h GSM calculations failed to converge numerically and displayed oscillations on the order of several tens of keV. Despite this, the energy differences compared to the 2p-2h truncation remained below 200~keV. These findings confirm that the 2p-2h truncation is sufficient to achieve converged results.
Additionally, we varied the leading-order parameter of the EFT interaction, which dominates within the EFT framework, by $2\%$. This variation resulted in ground state energy shifts of approximately 1 MeV in $^{18}$Mg and $^{19}$Al, and about 1.4 MeV in $^{20}$Si. Meanwhile, excitation energies and widths changed by less than 300~keV across all three nuclei.
Overall, these results indicate that the main source of uncertainty in results is the choice of EFT interaction parameters, while model space and truncation effects are minor.

Detailed discussions of $^{15}$F, $^{16}$Ne, $^{17}$Na and $^{18}$Mg can be found in our previous work~\cite{PhysRevC.111.014302}. For $^{17}$Na, direct measurements are still lacking; nevertheless, for the two lowest states our GSM results are consistent with empirical potential-model predictions~\cite{PhysRevC.87.044315,PhysRevC.82.027310}. 
To further explore nuclei at extreme proton excess, we extended the GSM calculations to 
$^{19}$Al and $^{20}$Si.
For $^{19}$Al, the low-lying states are predicted to be very broad resonances, indicating large decay widths. For $^{20}$Si, interpretable as a $^{14}$O core plus six valence protons, our GSM predicts a ground state unbound to six-proton emission with decay energy $E_{6p}= 10.152$ MeV and decay width $\Gamma = 371$~keV. Moreover, the first $2^+_1$ state is predicted at an excitation energy of 1.7 MeV. To the best of our knowledge, no prior theoretical or experimental results have been reported for $^{19}$Al and $^{20}$Si. The present predictions, therefore, provide quantitative benchmarks and guidance for future experimental investigations.

\begin{figure}[htbp]
    \includegraphics[width=8.6cm]{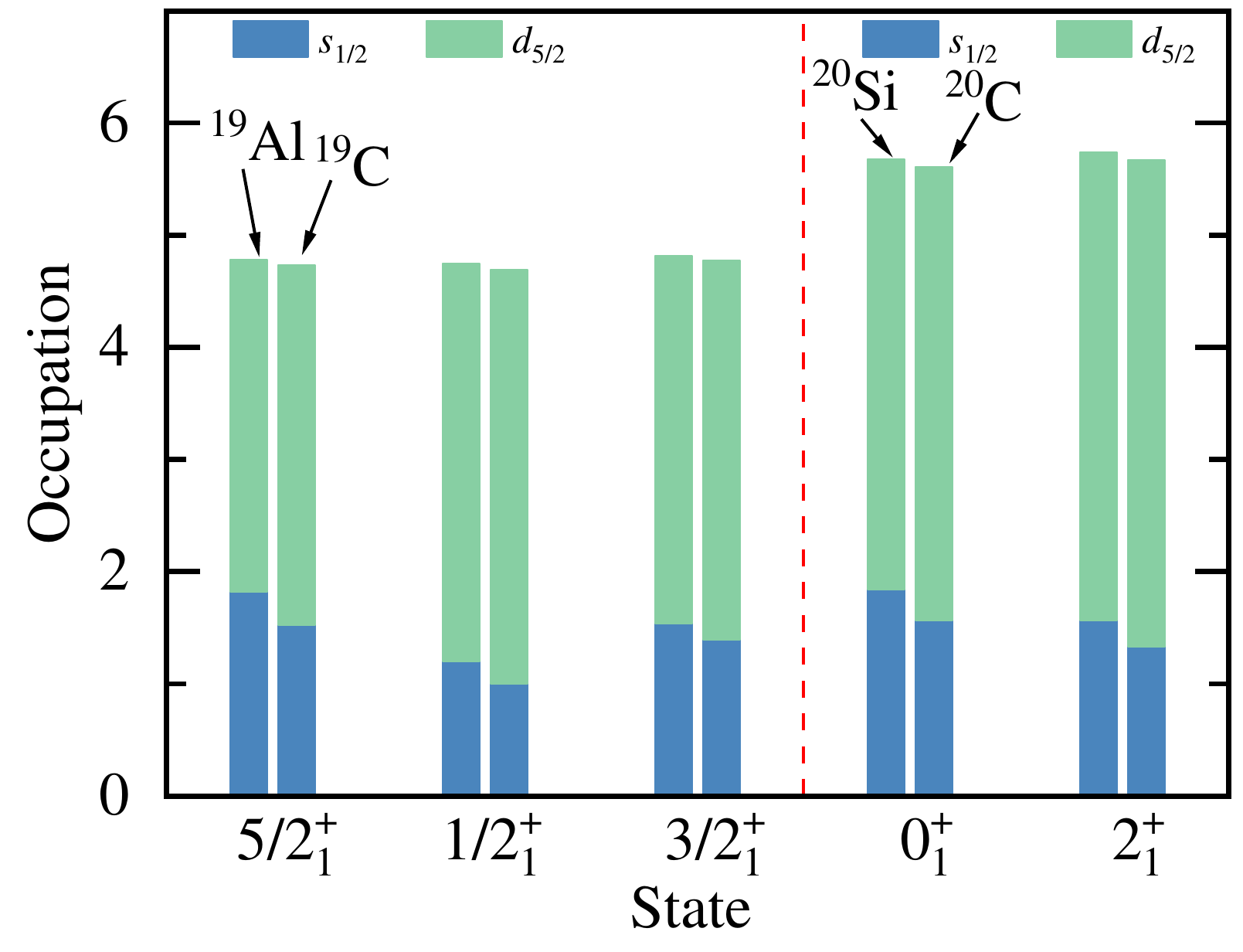}
    \caption{The calculated average occupations of the $s_{1/2}$ and $d_{5/2}$ partial waves for the low-lying states of $\rm{^{19}Al/ ^{19}C}$ and $\rm{^{20}Si/^{20}C}$.}
    \label{average occupations}
\end{figure}

The excitation energy of the first $2^+$ state, $E(2_1^+)$, of even-even nuclei serves as a sensitive probe for shell evolution~\cite{SORLIN2008602}. In $^{22}$O, the high $2^+_1$ energy confirms the magic number $N=14$~\cite{PhysRevLett.96.012501,THIROLF200016}. In contrast, $\gamma$-ray spectroscopy reveals that the $2^+_1$ energy in 
$^{20}$C is approximately half of that in $^{22}$O, indicating the disappearance of the $N=14$ subshell closure~\cite{PhysRevC.78.034315}. 
The nucleus $^{22}$Si has been predicted to be a double-magic with $Z=14$ and $N=8$, yet its predicted $E(2_1^+)$ is lower than in $^{22}$O, namely 2.18 MeV $\leq E(2_1^+) \leq 2.42$ MeV versus 3.2 MeV in $^{22}$O~\cite{LI2023138197}.
Recent precision mass measurements further support a $Z=14$ shell closure in $^{22}$Si~\cite{ffwt-n7yc}.
For $^{20}$Si, our GSM predicts $E(2_1^+)\approx 1.7$ MeV, which is smaller than the $^{22}$Si value from Ref.~\cite{LI2023138197} and comparable to that of $^{18}$Mg~\cite{PhysRevLett.127.262502}, suggests the absence of the $Z=14$ closed shell in $^{20}$Si. 
The underlying mechanism of the absence of shell closure can be attributed to continuum-induced deformation, whereby coupling to continuum reduces the energy gap between the $\pi s_{1/2}$ and $\pi d_{5/2}$ orbitals, enhancing their quadrupole coupling and driving significant deformation, which in turn obliterates the $Z=14$ shell closure. This scenario is consistent with findings in $^{22}$Al, where continuum coupling was shown to induce substantial deformation in the ground state within the particle-rotor model~\cite{PhysRevLett.132.152501}, as well as in GSM studies of neutron-rich fluorine isotopes~\cite{PhysRevC.106.034312}.
Furthermore, the mixed occupancies of the $s_{1/2}$ and $d_{5/2}$ partial waves, as shown in Table~\ref{table.1} and Fig.~\ref{average occupations}, support the picture of configuration mixing arising from such continuum-induced deformation.
The present GSM calculation includes only valence protons for $^{20}$Si. Although deformation is treated through configuration mixing within the shell model framework, a valence space restricted to protons alone may miss some non-negligible configurations that require explicit proton-neutron degrees of freedom for state-bearing deformation. As such, the quantitative results may be influenced by this limitation.

\begin{figure}[htbp]
    \centering
   \includegraphics[width=0.48\textwidth]{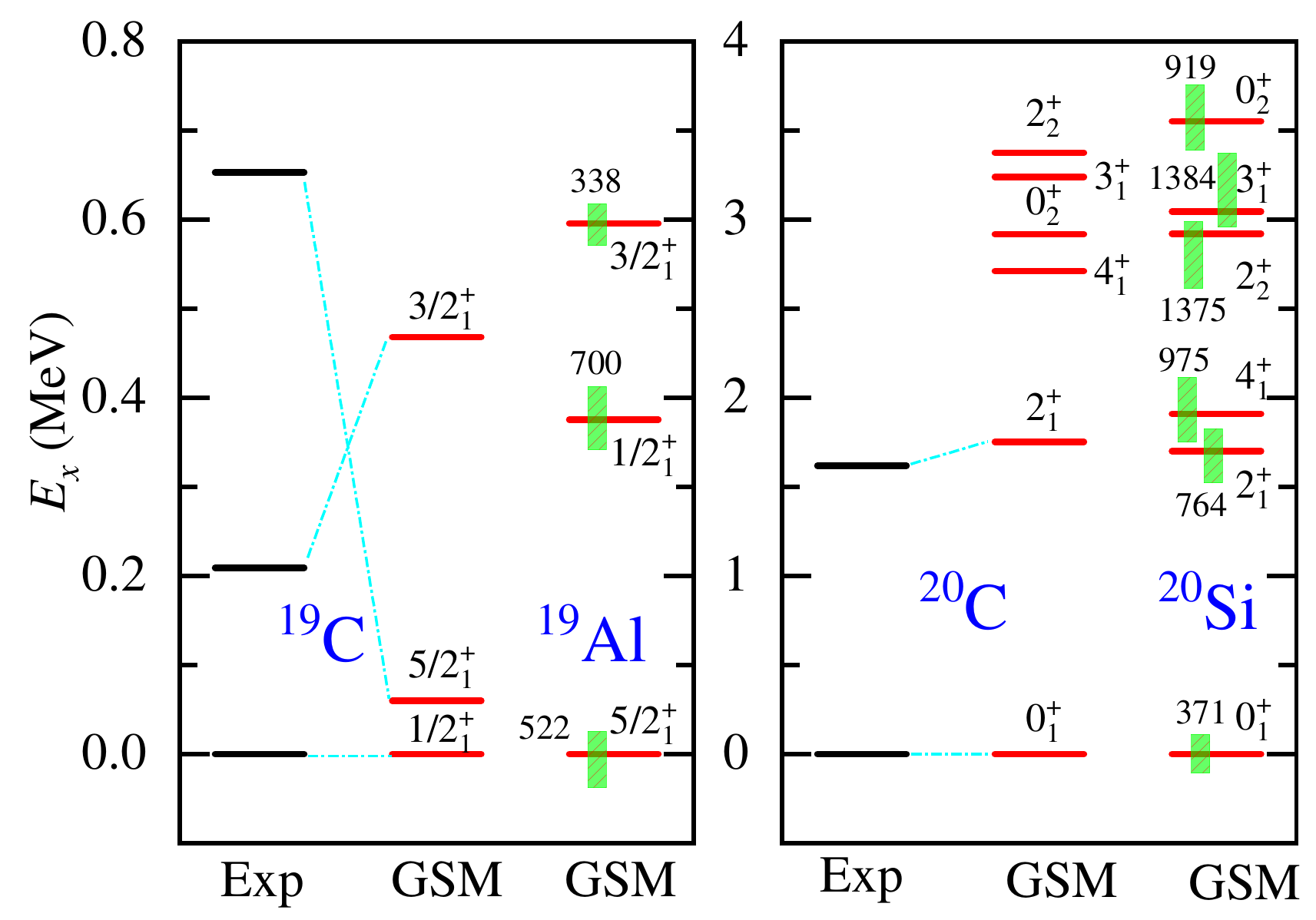}
    \caption{Comparison of the excitation energies and widths of mirroring nuclear states of $\rm{^{19}C/^{19}Al}$ and $\rm{^{20}C/^{20}Si}$. Experimental data of $\rm{^{19}C}$ and $\rm{^{20}C}$ are taken from Ref.~\cite{ensdf}.  }
    \label{relative energy}
\end{figure}

To further quantify isospin symmetry breaking in $^{19}$Al and $^{20}$Si, we computed the low-lying spectra of their mirror partners, $^{19}$C and $^{20}$C. The results are shown in Fig.~\ref{relative energy}. 
Both $^{19}$C and $^{20}$C are bound and well established experimentally~\cite{ensdf}.
Under exact isospin symmetry, mirror pairs would display the same level ordering and closely similar excitation energies. Thus, a comparison of the two spectra provides incisive constraints on the structure of the proton-rich mirrors $^{19}$Al and $^{20}$Si.

The three lowest states of $^{19}$C have been observed experimentally~\cite{THOENNESSEN20131}. 
Our GSM calculation reproduces these levels but predicts a reversal in the ordering of ${3/2}^+_1$ and the ${5/2}^+_1$ states. Moreover, GSM predicts a ground-state reversion in $^{19}\text{C}/^{19}\text{Al}$ mirror pair, as shown in the left panel of Fig.~\ref{relative energy}, the $^{19}$Al ground state is predicted to be ${5/2}^{+}$, whereas $^{19}$C has a $1/2^+_1$ ground state. A similar situation was reported recently for $^{20}$Al/$^{20}$N mirror pairs, combining new experimental data with GSM calculations, which established different ground-state spins for the two mirrors~\cite{hkmy-yfdk}.
For $^{20}$C, GSM calculations reproduce the low-lying states.

The mirror energy
difference (MED) for the $2^+_1$ states of $^{20}$Si/$^{20}$C mirror pair is nearly zero. Whereas, for the $4^+_1$, our calculation predicts that the excitation energies of $^{20}$Si are lower than those of $^{20}$C by approximately 800~keV. The unbound $s$ waves in the low-lying states of $^{19}$Al and $^{20}$Si should give a significant contribution to TES effects and exhibit significant isospin symmetry breaking. However, the phenomenon of TES is only found in the $1/2^+_1$ states of $^{19}$C/$^{19}$Al and the $4^+_1$ states of $^{20}$Si/$^{20}$C from the calculated low-lying spectra. Thus, demonstrating isospin symmetry breaking solely through the excitation energies of these mirror nuclei can be challenging.

Building on earlier demonstrations of dynamic TES, where differences in dominant many-body structures of mirror partners amplify TES effects~\cite{PhysRevLett.88.042502,PhysRevC.111.014302}, we therefore analyze, within GSM, the configuration mixing and average orbital occupations of the mirror pairs of $^{19}$Al/$^{19}$C and $^{20}$Si/$^{20}$C, as well as the contributions of various components of the GSM Hamiltonian.

\begin{table*}[htbp]
\caption{The main configurations and their corresponding probabilities of the low-lying excited states in $\rm{^{19}Al}/^{19}C$ calculated using GSM.
The tilde represents scattering continuum states. The imaginary part
of an operator’s expectation value in a resonant state can be interpreted as the uncertainty in measuring this expectation value~\cite{BERGGREN19961}. For $^{19}$C and $^{20}$C, the states calculated by GSM are all bound states, so the imaginary
parts of their corresponding probabilities are nearly zero and therefore omitted from the table.} 
\centering
\begin{tabular}{c@{\extracolsep{5pt}}ccccccccc}
\hline\hline
Nucleus & state & Configuration &Prob.& Nucleus & state &Configuration &Prob.\\ \hline
$^{19}$Al & $5/2^+_1$ & $(1s_{1/2})^{2}(0d_{5/2})^{3}$&$0.865 - i0.007$& $^{19}$C & $5/2^+_1$ &$(1s_{1/2})^{2}(0d_{5/2})^{3}$&$0.682$&\\
&&$(0d_{5/2})^{5}$&$0.036-i0.024$&&&$(0d_{5/2})^{5}$&$0.167$&\\
$^{19}$Al&${1/2}^{+}_{1}$ &$(1s_{1/2})^{1}(0d_{5/2})^{4}$&$0.763+i0.090$&$^{19}$C&$1/2^+_1$&$(1s_{1/2})^{1}(0d_{5/2})^{4}$&$0.814$
&\\ 
&&$(1s_{1/2})^{2}(0d_{5/2})^{2}(\widetilde{s_{1/2}})^{1}$&$0.107-i0.087$&&&$(1s_{1/2})^{1}(0d_{5/2})^{2}(0d_{3/2})^{2}$&$0.062$&\\
&&$(1s_{1/2})^{1}(0d_{5/2})^{2}(0d_{3/2})^{2}$&$0.023+i0.010$&&&$(1s_{1/2})^{1}(0d_{5/2})^{2}(\widetilde{f_{7/2}})^{2}$&$0.026$&\\
$^{19}$Al&${3/2}^{+}_{1}$ &$(1s_{1/2})^{2}(0d_{5/2})^{3}$&$0.524 + i0.042$& $^{19}$C&${3/2}^{+}_{1}$&$(1s_{1/2})^{1}(0d_{5/2})^{4}$&$0.475$
&\\
&&$(1s_{1/2})^{1}(0d_{5/2})^{4}$&$0.361 - i0.055$&&&$(1s_{1/2})^{2}(0d_{5/2})^{3}$&$0.386$\\

\hline
$^{20}$Si&$0^{+}_{1}$ &$(1s_{1/2})^{2}(0d_{5/2})^{4}$&$0.828+i0.002$& $^{20}$C& $0^{+}_{1}$ &$(1s_{1/2})^{2}(0d_{5/2})^{4}$&$0.667$
&\\ 
&&$(1s_{1/2})^{2}(0d_{5/2})^{2}(0d_{3/2})^{2}$&$0.065-i0.001$&&&$(0d_{5/2})^{6}$&$0.142$&\\ 
&&$(0d_{5/2})^{6}$&$0.035-i0.015$&&&$(1s_{1/2})^{2}(0d_{5/2})^{2}(0d_{3/2})^{2}$&$0.062$&\\
$^{20}$Si&$2^{+}_{1}$ &$(1s_{1/2})^{2}(0d_{5/2})^{4}$&$0.631+i0.204$&$^{20}$C&$2^{+}_{1}$ &$(1s_{1/2})^{1}(0d_{5/2})^{5}$&$0.506$
&\\
&&$(1s_{1/2})^{1}(0d_{5/2})^{5}$&$0.291-i0.173$&&&$(1s_{1/2})^{2}(0d_{5/2})^{4}$&$0.301$&\\
\hline\hline       
\end{tabular} 
\label{table.1}
\end{table*}

Dominant configurations of the valence nucleons above the inner core ($^{14}$O and $^{14}$C) for $^{19}$Al/$^{19}$C and $^{20}$Si/$^{20}$C are detailed in Table~\ref{table.1}, and Fig.~\ref{average occupations} displays the corresponding average occupations of the $s_{1/2}$ and $d_{5/2}$ partial waves. The occupation of other partial waves is minor and is not included in the figure.
For $^{19}$C, our GSM calculations give that the $5/2_1^+$ state is governed by $(1s_{1/2})^2(0d_{5/2})^3$ and  $(0d_{5/2})^5$ configurations with probabilities $68.2\%$ and $16.7\%$, respectively. In the mirror $^{19}$Al ground state ($5/2^+$), $(1s_{1/2})^2(0d_{5/2})^3$ configuration becomes over more dominant at $86.5\%$ (an increase of $18.3\%$ relative to $^{19}$C), while $(0d_{5/2})^5$ configuration is reduced to $3.6\%$. Consistently, Fig.~\ref{average occupations} shows a larger $s$-wave occupation for $5/2^+$ state in $^{19}$Al than in $^{19}$C. The absence of a centrifugal barrier for $s$ waves enhances the TES; together with the differing configuration mixes, this constitutes clear evidence for $dynamic$ TES in the $5/2_1^+$ mirror state in $^{19}$Al/$^{19}$C. Similar differences in many-body configurations and $s$-wave occupations are also observed in the $1/2^+_1$ and $3/2^+_1$ excited states, implying $dynamic$ TES in these states as well.

For $^{20}$Si/$^{20}$C (lower part of Table~\ref{table.1}; right panels of Fig.~\ref{average occupations}), both ground state are dominated by $(1s_{1/2})^2(0d_{5/2})^4$ configuration, yet the mirror partners differ in the detailed probability weights by about $16\%$, indicative of $dynamic$ TES already at the ground state. The $2_1^+$ states share the same principal components $(1s{1/2})^2(0d_{5/2})^4$ and $(1s_{1/2})^1(0d_{5/2})^5$, but with inverted amplitudes between the mirrors. Moreover, the $s$-wave occupations for the two lowest states in $^{20}$Si exceed those in $^{20}$C.
Overall, the obtained different many-body configurations and average occupations 
across the low-lying states  provide a coherent picture of $dynamic$ TES in $^{20}$Si/$^{20}$C, analogous to that identified in $^{19}$Al/$^{19}$C, as well as in $^{16}$Ne/$^{16}$C and $^{18}$Mg/$^{18}$C~\cite{PhysRevC.111.014302}.

\begin{figure}[htbp]
    \centering
    \includegraphics[width=8.6cm]{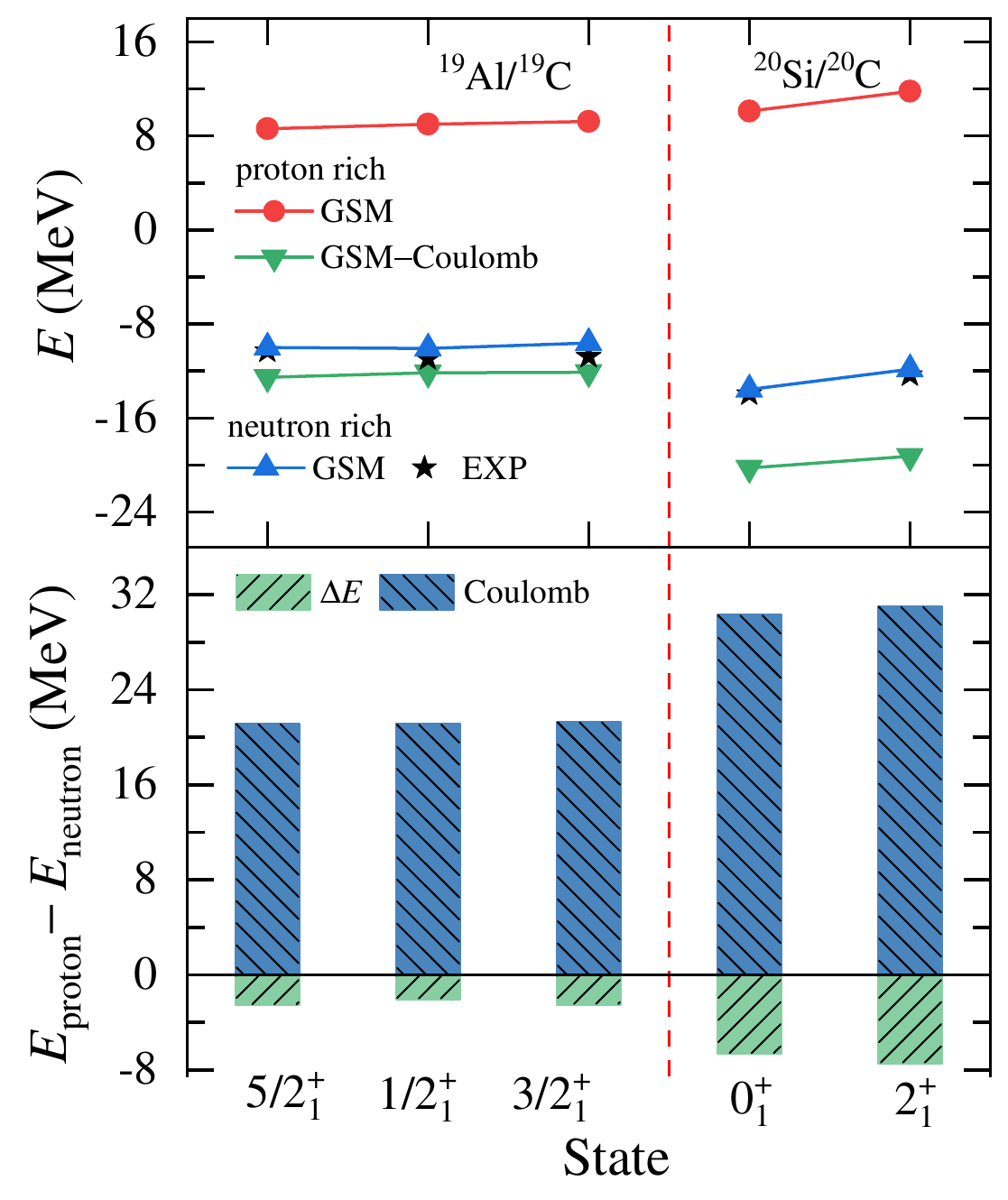}
    \caption{The contributions of different components of the Hamiltonian calculated by GSM to the low-lying excited states in $\rm{^{19}Al/^{19}C}$ (the left part) and $\rm{^{20}Si/^{20}C}$ (the right part). The
upper panel shows the absolute energies relative to their inner core. GSM-Coulomb represent the calculated energies
of $\rm{^{19}Al}$ or $\rm{^{20}Si}$ minus the Coulomb force contribution.
   The lower panel shows the value of each contribution, where $\Delta E$ is the
difference between contributions of nuclear forces in $^{19}$Al/$^{19}$C or $^{20}$Si/$^{20}$C.
Experimental data are taken from  Ref.~\cite{ensdf}.}
    \label{interactions}
\end{figure}

The GSM Hamiltonians for proton-rich nuclei and their neutron-rich mirror counterparts differ only in the Coulomb term. Thus, in the framework of isospin symmetry, energy differences between their mirror states should arise exclusively from Coulomb interactions.
However, previous studies have shown that both Coulomb and nuclear interactions contribute significantly to the energy differences of mirror states exhibiting $dynamic$ TES characteristics~\cite{PhysRevC.111.014302}. 
Consequently, to further investigate isospin symmetry breaking in the $^{19}$Al/$^{19}$C and $^{20}$Si/$^{20}$C mirror pairs from a Hamiltonian perspective, we separately calculated the contributions of the Coulomb and nuclear interactions. For the neutron-rich nuclei $^{19,20}$C, only the nuclear interaction was included in our GSM calculations, whereas for the proton-rich nuclei $^{19}$Al and $^{20}$Si, the nuclear interaction contributions were obtained by subtracting the Coulomb force contributions from the GSM total energy. The specific results and available experimental values are presented in Fig. \ref{interactions}, where $\Delta E$ denotes the differences in nuclear interaction contributions between the mirror states~\cite{PhysRevC.111.014302}.

Similarly to previous studies~\cite{PhysRevC.111.014302}, our GSM calculations reveal that both Coulomb and nuclear interactions make significant contributions to the energy differences between mirror states of $^{19}$Al/$^{19}$C and $^{20}$Si/$^{20}$C, with their respective contributions varying across states.
For the three lowest mirror states of $^{19}$Al/$^{19}$C, the Coulomb contributions are nearly uniform, with a maximum deviation of only 206~keV between the $1/2^+_1$ and $3/2^+_1$ states. Furthermore, all three mirror states exhibit large $\Delta E$ values exceeding 2 MeV, highlighting  pronounced isospin symmetry breaking in $^{19}$Al/$^{19}$C mirror pair.

Similar calculations were performed for the $0^+_1$ and $2^+_1$ mirror states in $^{20}$Si/$^{20}$C.
The Coulomb contribution to the energy difference in the $2^+_1$ states exceeds that in the $0^+_1$ states by approximately 545~keV. 
Furthermore, both mirror states exhibit more pronounced $\Delta E$ values, exceeding 6 MeV, than those in $^{19}$Al/$^{19}$C, indicating pronounced isospin symmetry breaking in $^{20}$Si/$^{20}$C as well.

These results combined with calculations of many-body configurations and average occupations provide compelling evidence for the presence of significant $dynamic$ TES characteristics in the $^{19}$Al/$^{19}$C and $^{20}$Si/$^{20}$C mirror nuclei. 
This finding underscores the complex interplay between Coulomb and nuclear interactions in exotic nuclei. 
Furthermore, our work validates the assertion of Ref.~\cite{PhysRevC.111.014302}: As a precise observable of dynamic TES, $\Delta E$ offers a new means to study isospin symmetry breaking in proton-rich nuclei near the drip line.

\section{Summary}

Within the valence nucleons plus core framework, $^{20}$Si, with a $^{14}$O core, is predicted to be the most promising candidate for 6$p$ decay. The GSM, which effectively treats both many-body correlations and continuum coupling, is the tool of choice to study such unbound nuclei. Thus, we performed GSM calculations of the low-lying spectra and particle decay widths for $^{20}$Si and carbon isotones bearing $A = 15\text{-}18$. As the available experimental values are well reproduced, the resulting GSM calculations are predictive. 
Our GSM predicts the $^{20}$Si ground state is unbound against six-proton emission, with decay energy $E_{6p}= 10.152$ MeV and a decay width of $\Gamma = 371$~keV. Furthermore, the first $2^+$ state is predicted to lie at approximately 1.7 MeV, a value comparable to that in $^{18}$Mg and lower than the \textit{ab initio} prediction for the $2_1^+$ state in $^{22}$Si, suggesting the disappearance of the $Z = 14$ subshell closure in $^{20}$Si.

To gain deeper insight into the structural properties of the low-lying states in $^{20}$Si and $^{19}$Al, we investigated isospin symmetry breaking in the mirror pairs $^{19}$Al/$^{19}$C and $^{20}$Si/$^{20}$C. Analyses of the calculated many-body configurations and average occupation provide strong evidence for pronounced $dynamic$ TES in these mirror systems. Moreover, we further investigated the isospin symmetry breaking from the perspective of the Hamiltonian by separately calculating the contributions of the Coulomb and nuclear interactions to the total nuclear energy. Our results indicate that the differences in the energy contributions from the nuclear interaction are significant and serve as sensitive probes of pronounced $dynamic$ TES.

Overall, the results reveal substantial energies and decay widths for both the ground and excited states of $^{20}$Si. The low-lying states of the mirror nuclei $^{19}$Al/$^{19}$C and $^{20}$Si/$^{20}$C exhibit clear signatures of a $dynamic$ TES, indicating complex interplay between Coulomb and nuclear interactions in these exotic nuclei. These results provide a robust theoretical foundation for future experimental and theoretical investigations.

\section{Acknowledgments}

This work has been supported by the National Key R\&D Program of China under Grant No. 2024YFE0109800, 2024YFE0109802, and 2023YFA1606403; the National Natural Science Foundation of China under Grant Nos. 12205340, 12175281, 12347106, 12575124, 12441506, and 12405141; the Gansu Natural Science Foundation under Grant Nos. 22JR5RA123 and 23JRRA614; the Natural Science Foundation of Henan Province No. 242300421048. The numerical calculations in this paper have been done at Hefei Advanced Computing Center.

\section*{References}

\bibliographystyle{elsarticle-num_noURL}
\bibliography{main}

\end{document}